# Thermodynamic- Kinetic Correlations in Supercooled Liquids- A Critical Survey of Experimental Data and Predictions of the Random First Order Transition Theory of Glasses

*Jacob D. Stevenson and Peter G. Wolynes\**

Department of Physics and Department of Chemistry and Biochemistry

University of California, San Diego

La Jolla, CA 92093

jstevens@physics.ucsd.edu

Thermodynamics and kinetics are thought to be linked in glass transitions. The quantitative predictions of $\alpha$-relaxation activation barriers provided by the theory of glasses based on random first order transitions are compared with experiment for 44 substances. The agreement found between the predicted activation energies near $T_g$ and experiment is excellent. These predictions depend on the configurational heat capacity change on vitrification and the entropy of melting the crystal which are experimental inputs.

The glass transition, as observed in the laboratory is a kinetic phenomenon. Unlike crystallization, the definition of the transition to the glassy state depends on experimental time scales. The change in mechanical response is accompanied by changes in thermal properties. Most dramatically the heat capacity drops upon cooling through the transition. This drop is quite measurable and generally appears to approach a discontinuity as the experimental time scale is increased. A variety of mean field models of disordered spin systems, electrical materials and molecular fluids predict a true thermodynamic transition with such a heat capacity discontinuity. In these models the thermodynamic transition occurs at a temperature $T_K$, where the configurational entropy of different mean field solutions vanishes[1-13]. These mean field frozen configurations first appear discontinuously at a higher dynamical transition temperature $T_A$, which coincides with the mode coupling transition for these models. It is natural to take these mean field theories, even given their status as approximations, as the starting point to understand the laboratory glass transitions, much as mean field theories are the natural starting point to understand critical phenomena[14,15] and nucleation dynamics at ordinary first order transitions[16,17]. But they are only a starting point. Indeed just as mean field theories of ordinary first order transitions must be supplemented by Maxwell's construction[18] and by a theory of nucleation rates[19,20] in order to describe how ordinary first order transformations occur in the laboratory, the connection of the existing mean field theories with real supercooled liquids can be made only by constructing "entropic droplets" which smear out the pure dynamical transition at $T_A$ and function as the mechanism of the slow α relaxation, which when it falls out of equilibrium, characterizes glass formation. If the underlying entropy crisis at $T_K$ remains in finite range systems, (a debatable point[21]) a dynamical theory based on entropic droplets yields a free energy barrier consistent with the commonly used Vogel-Fulcher law having a divergence at a temperature $T_0$, which according to this theory coincides with $T_K$:

$$\tau_\alpha = \tau_0 e^{\frac{DT_0}{T-T_0}}$$

While the well-known observed confluence of kinetic divergence at $T_0$ and the entropy crisis at $T_K$ is based on extrapolation from the experimentally accessible time scales, and has thus been questioned, this confluence is naturally explained by random first order transition (RFOT) theory. As Angell has often pointed out[22], the empirical correlation between thermodynamics and kinetics goes much further than merely this confluence of $T_0$ and $T_K$. The further connection can be seen in the coefficients D. The coefficient, D, varies considerably from substance to substance. The key experimental finding is that the larger is the observed heat capacity change, $\Delta C_p$, at $T_g$, the smaller D is. This has led to the concept of "fragility" of liquids which can either be termed "strong" or "fragile" depending on their D values. Within the RFOT framework, understanding the observed thermodynamic/kinetic correlations over a wide range of substances requires a microscopic theory of the free energy cost of entropic droplets. This microscopic theory was provided by Xia and Wolynes[23], who, using a density functional treatment of a glass transition of a fluid of spherical particles[24], argued that near $T_g$ this free energy cost depended both on the configurational entropy drive to form the droplets and on a free energy cost for mismatched areas, $\sigma$, which the density functional theory relates to the entropy cost of localizing the particles, $\sigma_0 = ¾ \, r_0^{-2} \, k_B T \log(\alpha/\pi e)$. The former term's dependence on temperature is reflected in $\Delta C_p$. The latter mismatch energy depends only logarithmically on the effective spring constant, $\alpha$, characterizing the caging. $\alpha$ is the inverse square amplitude of vibrational motions in a glassy configuration to the interparticle spacing and is related to the Lindeman ratio. The Lindemann ratio has been measured via neutron scattering and predicted by density functional theory. It varies only slightly from substance to substance, and $\alpha$ is found to be of order 100, the value we will use. Therefore the mismatch energy, which depends only logarithmically on this ratio, is predicted to be a nearly universal quantity in units of $k_B T_g$. In this way, as discussed in refs. 23 and 25, if the transition were an ordinary first order one the free energy of a nucleating droplet would be given as a function of the radius of the droplet, as:

$$F(r) = -\frac{4}{3}\pi T s_c r^3 + 4\pi\sigma r^2$$

The critical value of r from this theory gives a reconfiguration barrier proportional to $1/S_c^2$. For a random first order transition there is a multiplicity of solutions that can "wet" the droplet. To account for this, RFOT theory uses an idea from Villain first worked out for the random field Ising model to estimate how much the interface is wetted by specific solutions that better match the original. This wetting lowers the mismatch energy, $\sigma(r)=\sigma_0(r_0/r)^{1/2}$, and leads to the free energy barrier scaling with $1/S_c$, $\Delta F^{\ddagger}=3\pi\sigma_0^2 r_0/TS_c$. This wetting argument also restores consistency of the critical exponents at $T_K$ with hyperscaling. Combining the mismatch energy with the Xia-Wolynes value of $\sigma_0$ it immediately follows that D and $\Delta C_p$ should be inversely related, as was generally observed[26]. Furthermore the numerical coefficient of the mismatch energy is predicted by the microscopic calculation, crude as it is in some respects, so that the specific relation $D= 32k_B/\Delta C_p$ follows from the Xia-Wolynes treatment and can be tested. It is important that this relation is predicted for spherical particles and therefore the $\Delta C_p$ must refer to the heat capacity change for each of these spheres which might be called "beads". Glasses can be chemically complex. Many glasses are clearly mixtures of nearly spherical entities, such as $KCaNO_3$. In such cases counting "beads" is trivial. In other cases chemical intuition allows a reasonable mapping of the molecular shapes on to an aggregate of spherical objects e.g. o-terphenyl involves three fused benzene rings, so it can be thought of crudely as consisting of 3 "beads". With only modest ambiguity, structural chemical knowledge usually would allow the measured change in heat capacity per mole of many glass-forming substances to be converted to a heat capacity change per "bead". In this way Xia and Wolynes tested the predicted relation $D= 32k_B/\Delta C_p$ for 5 substances and the microscopically predicted correlation was shown to be reasonably accurate.

Wang and Angell[27] made a survey of 44 substances with an eye to establishing quantitative relations between their thermodynamic and kinetic properties on a purely empirical basis without involving any microscopic theory. Their analysis deftly avoids entirely the question of bead count. They found an excellent correlation between, on the kinetic side, the so-called m values of the liquids, characterizing their activation energies at $T_g$ and, on the thermodynamic side, $\Delta C_p$ (measured per mole), the glass transition temperature, $T_g$, and the latent heat of fusion per mole, $\Delta H_m$. The m value, is related to the D described above, in fact m is essentially the activation energy at the laboratory $T_g$ in units of $k_B T_g$

$$m = \left.\frac{\partial \log \tau}{\partial (T_g/T)}\right|_{T=T_g}$$

The empirical relation found by Wang and Angell is

$$m = 56\frac{T_g \Delta C_p}{\Delta H_m}$$

At first sight it appears strange that this relation should include the latent heat of freezing, since, after all, the freezing transition is by-passed, strictly speaking, when a liquid supercools.

Lubchenko and Wolynes[25] suggested a theoretical route to a correlation of this form by using the same density functional style argument to characterize crystallization as was used to characterize vitrification by Xia and Wolynes. If each "bead" of a molecular fluid becomes fully localized in a three dimensional sense in the crystal, then, the entropy of fusion per mole should be roughly the bead count per mole times the standard entropy of fusion of a spherical Lennard-Jones system (Clearly this is an approximate relation since the small density change on freezing depends on details of the attractive forces which will vary from substance to substance). Thus we can write

$$N_{bead} = \frac{\Delta H_m}{T_m}\frac{1}{S_{LJ}}$$

The entropy of fusion of Lennard-Jones spheres, per particle, $S_{LJ}$ is $1.68 k_B$[25]. When this relation is combined with the microscopic Xia-Wolynes prediction for $\Delta F^{\ddagger}$, and the well known form of $S_c$, $S_c = S_\infty(1-T_K/T)$ with $S_\infty$ given by $\Delta C_p(T_g)T_g/T_K$, one obtains the result

$$m = \frac{T_m}{\Delta H_m} \Delta C_p \left\{ \frac{32 S_{LJ} \log(e)}{S_c^2(T_g)} \right\}$$

The residual entropy per bead at $T_g$ is predicted by the Xia - Wolynes theory also to be universal[25] $S_c(T_g)=.82$. Thus Lubchenko and Wolynes predicted the correlation

$$m = 34.7 \frac{T_m}{\Delta H_m} \Delta C_p$$

While resembling Wang and Angell's correlation, strictly speaking, this prediction differs in form from the empirical correlation introduced by them by containing the melting temperature in addition to glass transition temperature. Yet, a commonly used empirical rule for simple substances is that $T_m=3/2\, T_g$. If this empirical relation is deemed to also hold, then Lubchenko and Wolynes pointed out that the Wang-Angell correlation is to be expected on the basis of the RFOT theory but with a theoretically predicted slope of 52, in contrast to the empirical slope of 56.

In this paper we check the predicted Lubchenko-Wolynes relation[25] directly without assuming $T_m/T_g \cong 3/2$. Thus no empirical relation between $T_m$ and $T_g$ is invoked in the present analysis. The melting characteristics come in only as a way of relating the entropy costs of localizing real molecules that are not spherical to their locations in the crystal and the entropy loss for localizing "beads" which are assumed to be spheres, as envisioned by Lubchenko and Wolynes. A list of the substances and their properties is provided in Table 1. The data were kindly provided by Angell to us and are supplementary material to the Wang-Angell paper[27]. Notice the effective bead counts differ from the nearest integer by typically 10-20%. This reflects the approximate nature of the mapping. Figure 1 plots the measured $m_{exp}$ versus the result predicted from the Lubchenko and Wolynes relation. The LW relation contains neither adjustable nor ambiguous quantities if the melting transition can be taken to be like that of the Lennard-Jones system. We see there is excellent agreement for the vast majority of the 44 substances for which we have all the relevant thermodynamic and kinetic data, for the freezing and vitrification transitions. Only for 8 substances is the error greater than 25%, while the dynamic range exhibited by m and $\Delta C_p$ is a factor of 5 or so. With the 8 outliers removed, the current correlation using $T_m$ rather than $T_g$ shows equally as tight a fit as that obtained by Wang and Angell with the same outliers missing; the R values of a best fit line are both 0.96.

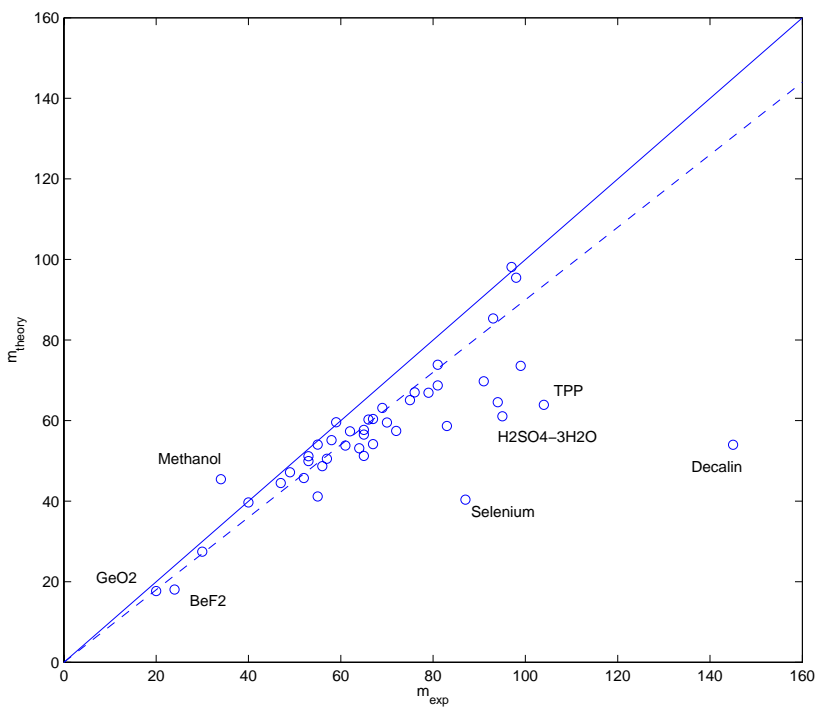

**Figure 1**. A plot of the fragility, m, measured from experiment vs. the theoretical estimate derived from the random first order transition theory. The solid line plots the perfect match, m(exp.) = m(theory), the dotted line, with slope .9, line gives the best fit. The experimental data used are in table 1 and are found in ref. 27.

Several aspects of Fig. 1 are worthy of more careful attention. First we note many of the extreme outliers are systems where the assumptions of the microscopic analysis are known to be violated in some way. Selenium, for example, is know to undergo a polymerization transition in the temperature range of the glass transition itself - clearly a changing degree of polymerization would violate the fixed, near-spherical unit assumption. Decalin, TPP and $H_2SO_4 3H_2O$ undergo crystallization transitions to orientationally disordered, i.e. plastic crystals, or exist as "glacial" liquid phases. In either case the assumption of complete freezing of degrees of freedom in the crystal or release of degrees of freedom on melting would be violated. Even if we were to remove these outliers it is clear that there are some modest but systematic deviations from the LW prediction. We must remember, however, that within the context of RFOT theory some such systematic deviation is to be expected[25,28]. The XW estimate of the mismatch energy assumed a maximally sharp interface between the mobilized region and its environment. Two effects within RFOT theory should broaden this interface and this would be expected to reduce the mismatch energy. One effect is that the proximity between $T_A$ and $T_G$ allows order parameter fluctuations to soften the barrier[25]. This is a temperature dependent effect. Also, in more elaborate theories, the details of replica symmetry breaking ("wetting[28]") in the interface can change the surface cost and introduce another length scale. Only a modest change of the mismatch energy estimate would be needed to bring any of the measured substances into perfect agreement with theory. It appears, however, that overall a small hardening is needed. We should, however, also keep in mind that not all of the substances surveyed are so well modeled with Lennard-Jones attractions, and thus they may have different density changes on freezing which would modify their entropy of fusion. This is a likely contribution to the deviation from theory.

This survey of experimental data on simple molecular substances leads us to conclude that the empirical correlations of kinetic and thermodynamic data for glasses and supercooled liquids summarized in Angell's notion of fragility are robust, at least for those systems credibly modeled as interacting (fused) spheres. Others have raised concerns about the earlier empirical correlations for polymeric systems[29]. Yet the necessary information about crystallization for these polymer systems is hard to monitor, making the present method of analysis difficult.

While we have observed some molecule specific deviations from the pattern of correlation between dynamics and thermodynamics, the observed correlation is extremely clear. Explaining these correlations, in our view and that of others, should be required of any microscopic theory of the glass transition in supercooled molecular liquids[30]. As we have shown here, the microscopic theory of glasses based on random first order transitions not only predicts these robust trends but provides quite accurate quantitative predictions of kinetic fragility from thermodynamics without the use of any adjustable fitting parameters for 44 substances.

We are grateful to M. Wang and C.A. Angell for helpful discussions and for making available to us the data they used for their previous analysis of kinetic-thermodynamic correlations in supercooled liquids that was listed as supplementary material in their paper[27]. This work was supported by NSF grant CHE0317017, "The Energy Landscapes of Glasses, Liquids, and Solutions".

**Table 1.** A survey of experimental data and theoretical results for 44 materials, including an explicit calculation of m(theory) predicted by RFOT and the number of beads per molecule for each substance, predicted from the melting entropy.

| Materials | Tg (K) | m (exp.) | m (theory) | Tm/Tg | ΔCp (J/Kmole) | ΔHm (kJ/mole) | Tm (K) | no. beads | Refs. |
|---|---|---|---|---|---|---|---|---|---|
| GeO2 | 810 | 20 | 17.7 | 1.71 | 6.27 | 17.1 | 1388 | 0.9 | 1 |
| BeF2 | 590 | 24 | 18.0 | 1.40 | 3 | 4.76 | 825 | 0.4 | 1,2 |
| ZnCl2 | 380 | 30 | 27.4 | 1.56 | 13.7 | 10.24 | 591 | 1.2 | 3 |
| methanol | 106 | 34 | 45.5 | 1.65 | 24.4 | 3.26 | 175 | 1.3 | 4,5,6 |
| n-propanol | 96.2 | 40 | 39.7 | 1.55 | 41.5 | 5.4 | 148.8 | 2.6 | 7,8 |
| Butyronitrile | 97 | 47 | 44.5 | 1.66 | 39.87 | 5.02 | 161.4 | 2.2 | 9 |
| Mg65Y10Cu25 | 402 | 49 | 47.1 | 1.82 | 16.1 | 8.65 | 730 | 0.8 | 10 |
| Ethylene glycol (ethane diol) | 152.7 | 52 | 45.7 | 1.71 | 60 | 11.86 | 260.5 | 3.3 | 11,8 |
| Ethanol | 97 | 55 | 41.2 | 1.64 | 37.3 | 5 | 159 | 2.3 | 12,8 |
| m-xylene | 125.5 | 56 | 48.7 | 1.80 | 72 | 11.56 | 225.3 | 3.7 | 13 |
| Glycerol | 190 | 53 | 49.9 | 1.53 | 90.5 | 18.3 | 291 | 4.5 | 14 |
| 3-bromopentane | 107 | 53 | 51.1 | 1.56 | 74 | 8.4 | 167.3 | 3.6 | 8,15,11 |
| Pd40Ni40P20 | 570 | 55 | 54.0 | 1.55 | 16.54 | 9.39 | 884 | 0.8 | 16,17 |
| m-cresol | 198.6 | 57 | 50.5 | 1.44 | 54 | 10.57 | 285 | 2.7 | 13 |
| 2-methylpentane | 78 | 58 | 55.2 | 1.53 | 83.4 | 6.26 | 119.3 | 3.8 | 18 |
| Toluene | 117 | 59 | 59.6 | 1.52 | 64 | 6.64 | 178.15 | 2.7 | 19,8 |
| fructose | 286 | 61 | 53.8 | 1.32 | 133 | 32.43 | 378 | 6.1 | 20,21,22 |
| Phenolphthalein | 363 | 62 | 57.3 | 1.47 | 146 | 47.15 | 533.7 | 6.3 | 21,23 |
| indomethacin | 315 | 64 | 53.1 | 1.38 | 139 | 39.4 | 434 | 6.5 | 24 |
| 2-methyltetrahydrofuran | 91 | 65 | 51.2 | 1.53 | 72 | 6.77 | 138.8 | 3.5 | 15,8,1,25,26 |
| Hydrochloro-thiazide | 385 | 65 | 56.5 | 1.42 | 92.3 | 31 | 547 | 4.1 | 27,28,29 |
| griseofulvin | 364 | 65 | 57.7 | 1.36 | 127 | 37.75 | 494 | 5.5 | 27,28,29 |
| trinaphthylbenzene | 337 | 66 | 60.3 | 1.41 | 122 | 33.3 | 474 | 5.0 | 30,31,32 |
| di-2-ethylhexylphthalate | 187 | 67 | 60.4 | 1.44 | 116 | 18 | 270 | 4.8 | 33 |
| probucol | 295 | 67 | 54.2 | 1.35 | 139.5 | 35.66 | 399 | 6.4 | 27 |
| 9-bromophenanthrene | 224.75 | 69 | 63.1 | 1.48 | 76.5 | 14 | 333 | 3.0 | 34 |
| phenobarbital | 319 | 70 | 59.5 | 1.40 | 106.8 | 27.9 | 448 | 4.5 | 27,28,29 |
| D-glucose | 306 | 72 | 57.4 | 1.37 | 128 | 32.4 | 419 | 5.5 | 6,3,20,22 |
| glibenclamide | 331 | 75 | 65.1 | 1.36 | 222.3 | 53.35 | 450 | 8.5 | 27,28,29 |
| maltitol | 311 | 94 | 64.5 | 1.35 | 243.6 | 55 | 420 | 9.4 | 35,36,37 |
| Salol | 220 | 76 | 67.0 | 1.43 | 118.3 | 19.3 | 315 | 4.4 | 14 |
| m-toluidine | 187 | 79 | 66.9 | 1.33 | 68 | 8.8 | 249.5 | 2.5 | 13,38,3,39 |
| OTP | 246 | 81 | 73.9 | 1.34 | 111.27 | 17.2 | 329 | 3.7 | 14 |
| flopropione | 335 | 81 | 68.7 | 1.35 | 127.5 | 29.1 | 452 | 4.6 | 27,28,29 |
| α-phenyl-cresol | 220 | 83 | 58.7 | 1.49 | 120 | 23.3 | 328.2 | 5.1 | 40 |
| Selenium | 308 | 87 | 40.4 | 1.60 | 14.4 | 6.12 | 494.33 | 0.9 | 8,2 |
| triphenylethene | 248 | 91 | 69.8 | 1.38 | 120 | 20.35 | 341 | 4.3 | 33,41,42 |
| Sorbitol (D-glucitol) | 266 | 93 | 85.3 | 1.44 | 189.45 | 29.5 | 383 | 5.5 | 3,1,36,43,24 |
| H2SO4-3H2O | 158 | 95 | 61.1 | 1.50 | 180 | 24.22 | 236.8 | 7.3 | 33,2,44 |
| sucrose | 323 | 97 | 98.1 | 1.44 | 250 | 41.1 | 465 | 6.3 | 20,21,24 |
| Ca(NO3)2-4H2O | 217 | 98 | 95.5 | 1.46 | 270.5 | 31.17 | 317 | 7.0 | 2 |
| Propylene Carbonate (PC) | 159.54 | 99 | 73.6 | 1.37 | 75.4 | 7.77 | 218.6 | 2.5 | 45,46 |
| TPP | 200 | 104 | 63.9 | 1.49 | 155 | 25 | 297 | 6.0 | 24,21 |
| decalin | 138.42 | 145 | 54.0 | 1.66 | 64 | 9.46 | 230 | 2.9 | 47 |

**References for Table 1:**

(1) Böhmer, R.; Ngai, K.L.; Angell, C.A.; and Plazek, D.J. *J. Chem. Phys.* **1993,** 99(5), 4201

(2) Brawer, S. *J. Chem. Phys.* **1985,** 81(2), 954

(3) Angell, C.A. *J. Res. NIST* **1997,** 102(2), 183

(4) Karolczak, M. and Mohilner, D.M. *J. Phys. Chem.* **1982,** 86(15), 3845

(5) Smith, D.L. dissertation, Arizona State University, **1981**

(6) Naito, K.; and Miura, A. *J. Phys. Chem.* **1993,** 97(23), 6240

(7) Takahara, S.; Yamamuro, O.; and Suga, H. *JNCS* **1994,** 171(3), 259

(8) Moynihan, C.T.; and Angell, C.A. *JNCS* **2000,** 274, 131

(9) Hikawa, H.; Oguni, M.; and Suga, H. *JNCS* **1988,** 101, 90

(10) Busch, R.; Liu, W.; and Johnson, W.L. *J. Appl. Phys.* **1998,** 83, 4134

(11) Takeda, K.; Yamamuro, O.; Tsukushi, I.; Matsuo, T.; and Suga, H. *J. Molecul. Struct.* **1994,** 479, 227

(12) Haida, O.; Suga, H.; and Seki, S. *J. Chem. Thermo.* **1977,** 9(12), 1133

(13) Alba-Simionesco, C.; Fan, J.; and Angell, C.A. *J. Chem. Phys.* **1999,** 110(11), 5262

(14) Wang, L.M.; Velikov, V.; and Angell, C.A. *J. Chem. Phys.* **2002,** 117, 10184

(15) Richert, R.; and Angell, C.A. *J. Chem. Phys.* **1998,** 108(21), 9016

(16) Wilde, G.; Görler, G.P.; Willnecker, R.; and Dietz, G. *App. Phys. Lett.* **1994,** 65, 397

(17) Kawamura, Y.; and Inoue, A. *App. Phys. Lett.* **2000,** 77, 1114

(18) Douslin, D.R.; and Huffman, H.M. *J. Am. Chem. Soc.* **1946,** 68(9), 1704

(19) Yamamuro, O.; Tsukushi, I.; Lindqvist, A.; Takahara, S.; Ishikawa, M.; and Matsuo, T. *J. Phys. Chem. B* **1998,** 102(9), 1605

(20) Raemy, A.; Schweizer, T.F. *J. Ther. Ana.* **1983,** 28, 95

(21) J. Fan ASU dissertation

(22) Gangasharan, and Murthy, S.S.N. *J. Phys. Chem.* **1995,** 99(32), 12349

(23) Eduardo M. Sanchez ASU dissertation

(24) Shamblin, S.L.; Tang, X.; Chang, L.; Hancock, B.C.; and Pikal, M.J. *J. Phys. Chem. B* **1999,** 103(20), 4113

(25) Johari, G.P. Chem. Phys. **2001,** 265(2), 217

(26) Mizukami, M.; Fujimori, H.; and Oguni, M *Progress of Theoretical Physics Suppl.* **1997,** 126, 79

(27) Crowley, K.J.; and Zografi, G. Thermochim. *ACTA* **2001,** 380(2), 79

(28) Hancock, B.C.; and Parks, M. *Pharm. Res.* **2000,** 17, 397

(29) Aso, Y.; Yoshioka, S.; and Kojima, S. *J. Pharm. Sci.* **2000,** 89, 408

(30) Plazek, D.J.; and Magill, J.H. *J. Chem. Phys.* **1966,** 45(8), 3038

(31) Magill, J.H. *J. Chem. Phys.* **1967,** 47(8), 2802

(32) Tsukushi, I.; Yamamuro, O.; Ohta, T.; Matsuo, T.; Nakano, H.; and Shirota, Y. *J. Phys.: Cond. Matt.* **1996,** 8(3), 245

(33) Johari, G.P. *J. Chem. Phys.* **2000,** 113, 751

(34) Takahara, S.; Yamamuro, O.; and Matsuo, T. *J. Phys. Chem.* **1995,** 99(23), 9589

(35) Lebrun, N.; and van Miltenburg, J.C. *J. Alloys. Compounds.* **2001,** 320, 320

(36) Faivre, A.; Niquet, G.; Maglione, M.; et al. *Euro. Phys. Jour. B* **1999,** 10, 277

(37) Carpentier, L.; Bourgeois, L.; and Descamps, M. *J. Ther. Ana. and Calo.* **2002,** 68, 727

(38) Mbaze Meva'a, L.; and Lichanot, A. *Thermochimica ACTA* **1990,** 158, 335

(39) Cutroni, M.; Mandanici, A.; Spanoudaki, A.; and Pelster, R. *J. Chem. Phys* **2001,** 114(16), 7118

(40) Murthy, S.S.N.; Paikaray, A.; and Arya, N. *J. Chem. Phys.* **1995,** 102(20), 8213

(41) Fujimori, H.; and Oguni, M. *J. Chem. Thermodyn.* **1998,** 30, 509

(42) Hikima, T.; Hanaya, M.; and Oguni, M. *J. Molecul. Struct.* **1999,** 479, 245

(43) Nozaki, R.; Suzuki, D.; Ozawa, S.; and Shiozaki, Y. *JNCS* **1998,** 235, 393

(44) Kunzler, J.E.; Giauque, W.F. *J. Am. Chem. Soc.* **1952,** 74, 797

(45) Fujimori, H.; and Oguni, M. *J Chem. Thermodyn.* **1994,** 26, 367

(46) Angell, C.A.; Boehm, L.; Oguni, M.; and Smith, D.L. *J. Molecul. Liq.* **1993,** 56, 275

(47) Jackson, C.L.; and McKenna, G.B. *J. Chem. Phys.* **1990,** 93(12), 9002

**References:**

(1) Kirkpatrick, T.R.; and Wolynes, P.G. *Phys. Rev. A* **1987,** 35, 3072-3080

(2) Kirkpatrick, T.R.; and Thirumalai, D. *Phys. Rev. Lett.* **1987,** 58, 2091-2094

(3) Kirkpatrick, T.R.; and Wolynes, P.G. *Phys. Rev. B* **1987,** 36, 8552-8564

(4) Kirkpatrick, T.R.; Thirumalai, D.; and Wolynes, P.G. *Phys. Rev. A* **1989,** 40, 1045-1054

(5) Cugliandolo, L.F.; Kurchan, J.; Monasson, R.; and Parisi, G. *J. Phys. A* **1996,** 29, 1347-1358

(6) Mezard, M.; and Parisi, G. *Phys. Rev. Lett.* **1999,** 82, 747-750

(7) Gross, D.J.; Kanter, I.; and Sompolinsky, H. *Phys. Rev. Lett.* **1985,** 55, 304

(8) Gardner, E. *Nucl. Phys. B* **1985,** 257(6), 747-765

(9) Mezard, M.; Parisi, G.; and Virasuro, M.A. **1987** *Spin Glass Theory And Beyond* (World Scientific, Singapore)

(10) Biroli, G.; and Mezard, M. *Phys. Rev. Lett.* **2002**, 88, 25501

(11) Ciamarra, M.P.; Tarzia, M.; de Candia, A.; Coniglio, A. *Phys. Rev. E* **2003**, 67, 57105

(12) Lopatin, A.V.; Ioffe, L.B. *Phys. Rev. B* **2002**, 66, 174202

(13) Schmalian, J.; and Wolynes, P.G. *Phys. Rev. Lett.* **2000**, 85 (4), 836

(14) Anderson, P.W. "Basic Notions of Condensed Matter Physics," **1984** Westview Press

(15) Chaikin, P.M.; and Lubensky, T.C. *"Principles of Condensed Matter Physics,"* **1995** Cambridge University Press

(16) van Kampen, N.G. *Phys. Rev.* **1964,** 135, A362

(17) Langer, J.S. *Ann. Phys.* **1967,** 41(1), 108

(18) Maxwell, J.C. "*On the Dynamical Evidence of the Molecular Constitution of Bodies*," *Nature* **1875,** 11, 357-359, 374-377

(19) Sarkies, K.W.; and Frankel, N.E. *J. Chem. Phys.* **1971,** 54, 433

(20) Zel'dovich, J.B. *Zh. Eksp. Teor. Fiz.* **1942,** 12, 525

(21) Eastwood, M.P.; and Wolynes, P.G. *Europhys. Lett.* **2002,** 60(4), 587-593

(22) Angell, C.A. *J. Res. NIST* **1997**, 102 (2), 171

(23) Xia, X.Y.; and Wolynes, P.G. *PNAS* **2000,** 97(7), 2990-2994

(24) Singh, Y.; Stoessel, J.; and Wolynes, P.G. *Phys. Rev. Lett.* **1985,** 54, 1059-1062

(25) Lubchenko, V.; and Wolynes, P.G. *J. Chem. Phys.* **2002,** 119(17), 9088-9105

(26) Angell, C.A. *Science* **1995,** 267(5206), 1924-1935

(27) Wang, L.M.; and Angell, C.A. *J. Chem. Phys.* **2003,** 118, 10353

(28) Dzero, M.; Schmalian, J.; and Wolynes, P.G. *arXiv: Cond-Mat* 0502011 v1 1 feb 2005

(29) Huang, D.H.; and McKenna, G.B. *J. Chem. Phys.* **2001**, 114(13), 5621

(30) Biroli, G.; Bouchaud, J.P.; and Tarjus, G. *arXiv: cond-mat*/0412024 v1 1 Dec. 2004